# Magnetization dynamics in the density matrix formalism

Benjamin Assouline, Amir Capua*

Department of Applied Physics, The Hebrew University of Jerusalem, Jerusalem 91904, Israel

*e-mail: amir.capua@mail.huji.ac.il

**Abstract:**

**Magnetization dynamics described by the Landau-Lifshitz-Gilbert-Slonczewski (LLGS) equation can be formulated to have the form of the well-known two-level-system (TLS) equations. Recently, we showed that a DC spin-transfer torque (STT) term in the LLGS equation can be mapped to a modulation of the carrier relaxation rates in the analogous TLS equations. Here, we extend the analogy to the TLS dynamics by including the AC magnetic field, AC demagnetization field, and AC STT excitation that we show constitute the interaction term in the analogous TLS picture. Interestingly, we find that the carrier injection rate in the TLS equations that is responsible for transitions between the excited and ground states of the system naturally translates to an intense short magnetic pulse that reverses the magnetization state. Furthermore, we also show that the two helicities of circularly polarized magnetic pulses correspond to the two carrier injection rates in the analogous TLS picture. In the context of the highly debated all-optical helicity dependent switching experiment, it offers a new explanation of the magnetization reversal from first principles.**

**Introduction**

Two – level quantum systems (TLS) are extensively used to describe a variety of physical phenomena. Their main asset is their simplicity, as they represent the simplest dynamics of a quantum system. For example, the Bloch vector representation of a quantum state is used in quantum computing in the context of pi and half-pi pulses. TLS Hamiltonians are used in Topological insulator theory to describe energy band crossings, which contribute significantly to important quantities such as the Berry curvature. Regarding the domain of magnetism, it was early on realized by Feynman et al [10] that a magnetic moment precessing in equilibrium can be formulated as a TLS in steady population state. However, this derivation didn't account for magnetic losses, which are prominent in most realistic magnetic systems. In ref. [11], which accounts for magnetic losses, the TLS model was used to interpret a switching process of the magnetization in terms of population transfer of a TLS. Following ref. [11], in this paper we present a step by step mathematical derivation of the analogy between the LLGS and TLS dynamics, which accounts for magnetic damping, external and demagnetization field, and AC and DC spin-transfer torque (STT) excitation. We first discussed this model in the context of amplification of spin currents [12], in which the effect of DC STT on the analogues TLS was examined. It was found to modify the decay terms rather than induce a pump term, like the pump current in the semiconductor optical amplifier. It led us to ask which excitation in the LLGS would be equivalent to such a pump term in the TLS. We demonstrate that such an excitation is an oscillatory field with a Lorentzian envelope. We demonstrate the ability of such a field to switch the magnetization state, in a way reminiscent of the all-optical switching experiment.

## Results

We start with the Landau–Lifshitz–Slonczewski (LLS) representation of the LLGS equation [1]:

$$\frac{d\vec{M}}{dt} = -\frac{\gamma}{1+\alpha^2}\vec{M}\times\vec{H}_{eff} - \frac{\gamma\alpha}{1+\alpha^2}\frac{1}{M_s}\vec{M}\times\vec{M}\times\vec{H}_{eff} + \gamma\frac{H_{SHE}}{M_s}\vec{M}\times\vec{M}\times\vec{S} \quad (1)$$

where $\vec{M}$ is the magnetization vector, $M_s$ is the magnetization saturation, $\gamma$ is the gyromagnetic ratio, $\alpha$ is the Gilbert damping, $\vec{H}_{eff}$ is the effective magnetic field vector, $\vec{S}$ is the spin polarization vector of the STT excitation and $H_{SHE}$ is the spin Hall effect field term. We define the generalized magnetic fields $\vec{B}'$ and $\vec{H}'_{eff}$:

$$\vec{B}' \equiv \frac{\gamma}{1+\alpha^2}\vec{H}'_{eff} \equiv \frac{\gamma}{1+\alpha^2}\left(\vec{H}_{eff} - \frac{\alpha}{M_s}\vec{H}_{eff}\times\vec{M} + \frac{H_{SHE}\cdot(1+\alpha^2)}{M_s}\vec{S}_{AC}\times\vec{M}\right) \quad (2)$$

Following ref. [11], we define the quantities $\kappa$ and $\kappa_0$:

$$\begin{cases} \kappa \equiv \frac{1}{2}\left(B'_x - jB'_y\right) \\ \kappa_0 \equiv B'_z \end{cases} \quad (3)$$

where $\kappa$ represents the generalized AC excitation and $\kappa_0$ represents the generalized DC field. The equations of motion of TLS density matrix elements, using $\kappa$ and $\kappa_0$ as coefficients (see full derivation in Appendix 1):

$$\begin{cases} \dot{\rho}_{11} = j(\kappa^*\rho_{12} - \kappa\rho_{21}) = -j\kappa\rho_{21} + c.c. \\ \dot{\rho}_{22} = -j(\kappa^*\rho_{12} - \kappa\rho_{21}) = j\kappa\rho_{21} + c.c. \\ \dot{\rho}_{12} = -j\kappa_0\rho_{12} + j\kappa(\rho_{11} - \rho_{22}) \end{cases} \quad (4)$$

The generic equations of motion for the TLS density matrix coefficient describing the semiconductor laser can be written as [2-4]:

$$\begin{cases} \dot{\rho}_{11} = -\Lambda_h - \gamma_v\rho_{11} + \frac{j}{2}[(\rho_{12}-\rho_{21})(V_{12}+V_{21}) - (\rho_{12}+\rho_{21})(V_{12}-V_{21})] \\ \dot{\rho}_{22} = \Lambda_e - \gamma_c\rho_{22} - \frac{j}{2}[(\rho_{12}-\rho_{21})(V_{12}+V_{21}) - (\rho_{12}+\rho_{21})(V_{12}-V_{21})] \\ \dot{\rho}_{12} = -(j\omega + \gamma_{inh})\rho_{12} + j(\rho_{11}-\rho_{22})V_{12} \end{cases} \quad (5)$$

where $\Lambda_h$ ($\Lambda_e$) is the incoherent injection rate of holes (electrons) into the valence (conductance) band. $\gamma_v$ ($\gamma_c$) is the decay rate of the holes (electrons) in the valence (conductance) band, and $\gamma_{inh}$ is the rate of decay caused by the inhomogeneous broadening of the system. $\omega$ is the resonance frequency and $V_{12}$ is the interaction term. We apply the following transformation on the magnetization components from Eq. 1 and transform them to density matrix elements:

$$\begin{cases} M_x \equiv A_1 A_2^* + A_1^* A_2 = \rho_{12} + \rho_{21} = 2\Re(\rho_{12}) \\ M_y \equiv j(A_1 A_2^* - A_1^* A_2) = j(\rho_{12} - \rho_{21}) = -2\Im(\rho_{12}) \\ M_z \equiv |A_1|^2 - |A_2|^2 = \rho_{11} - \rho_{22} \\ M_s \equiv |A_1|^2 + |A_2|^2 = \rho_{11} + \rho_{22} \end{cases} \quad (6)$$

Under this transformation the magnetization represents the Bloch vector of an effective TLS. We substitute the expression for $\vec{M}$ (Eq. 6) in Eq. 4 and Eq. 5 and compare the two:

$$\begin{cases} -\Lambda_h - \gamma_v \rho_{11} + [M_y \Re\{V_{12}\} + M_x \Im\{V_{12}\}] \equiv -j\kappa\rho_{21} + c.c. \\ \Lambda_e - \gamma_c \rho_{22} - [M_y \Re\{V_{12}\} + M_x \Im\{V_{12}\}] \equiv j\kappa\rho_{21} + c.c. \\ -(j\omega + \gamma_{inh})\rho_{12} + jM_z V_{12} \equiv -j\kappa_0 \rho_{12} + j\kappa M_z \end{cases} \quad (7)$$

The right-hand side of Eq. 7 contains the terms $\kappa$ and $\kappa_0$ which consist of the excitations driving $\vec{M}$ in the LLGS equation, while the left-hand side contains terms that phenomenologically describe the TLS. This will allow us later on to express the TLS terms using the LLGS equation.

In order to drive the magnetization with a field that is equivalent to an incoherent injection pump in the analogues effective TLS (i.e. the $\Lambda_{h\backslash e}$ terms in Eq.7), we define $\vec{B}' \equiv \vec{B}_{pump[h]}$, and observe the requirement on $\vec{B}_{pump[h]}$ imposed from its corresponding $\kappa_{pump[h]}$ in Eq. 7:

$$\kappa_{pump[h]} = -\Lambda_h \frac{j}{2\rho_{21}} = -\Lambda_h \frac{j}{M_x + jM_y} = -\Lambda_h \frac{j(M_x - jM_y)}{M_x^2 + M_y^2} = -\Lambda_h \frac{jM_x + M_y}{M_s^2 - M_z^2} \quad (8)$$

where we have used the normalization condition of $\vec{M}$ (i.e. $M_x^2 + M_y^2 + M_z^2 = M_s^2$). By defining $\Lambda_h \equiv -\frac{\gamma}{2(1+\alpha^2)} \Lambda_p$, and comparing the real and imaginary parts on both sides, we obtain the following effective pump field:

$$\vec{H}_{pump[h]} = \frac{\Lambda_p}{M_s^2 - M_z^2} \begin{pmatrix} M_y \\ -M_x \\ 0 \end{pmatrix} \quad (9)$$

The incoherent injection terms have opposite signs in their respective rate equations, so that the equivalent pump field for electron injection, $\vec{H}_{pump[e]}$, has the opposite sign of $\vec{H}_{pump[h]}$:

$$\vec{H}_{pump[e]} = \frac{\Lambda_p}{M_s^2 - M_z^2} \begin{pmatrix} -M_y \\ M_x \\ 0 \end{pmatrix} \quad (10)$$

The $\vec{H}_{pump}$ fields in Eq. 9, 10 are oscillatory field whose components are rotated by $\pm 90°$ to $M_x$ and $M_y$. Their magnitude is proportional to a Lorentzian envelope $1/(M_s^2 - M_z^2)$, which is related to the population inversion of the TLS. Near the poles (i.e. $|M_z| \approx M_s$) the $\vec{H}_{pump}$ field's magnitude diverges because of the $1/(M_s^2 - M_z^2)$ term, which corresponds to conservation of angular momentum, as it would require an infinite torque to pump the system more than its saturation value $M_s$ (i.e. as a Bloch vector, $\vec{M}$ satisfies $|\vec{M}|^2 = M_s^2$). A key difference between the presented $\vec{H}_{pump}$ fields and the pump process in the optical amplifier is that, in the latter, pumping of the system through incoherent injection is realized without phase information. The phase of the incoherent injected charge carries is locked to that of the system during the first few cycles of the interaction. On the other hand, the $\vec{H}_{pump}$ fields in Eq. 9, 10 require phase information of the system because of their explicit dependence on $M_x$ and $M_y$. In particular, $\vec{H}_{pump[e]} = -\vec{H}_{pump[h]}$, so that under a phase shift of half a cycle (i.e. 180°), $\vec{H}_{pump}$ will change its functionality - from pumping in a certain direction ($-\hat{z}$ for $\vec{H}_{pump[e]}$), into pumping in the opposite direction ($\hat{z}$ for $\vec{H}_{pump[h]}$). Interestingly, this behavior of the $\vec{H}_{pump}$ fields is reminiscent to that of the L and R polarized pulses used in all-optical switching experiments [5-8] since the L and R pulses are polarized 180° relative to each other, and switch $\vec{M}$ in opposite directions. The general form of $\vec{H}_{pump}$, which we denote by $\vec{H}_{pump\,\beta}$, is an oscillatory field with a Lorentzian envelope whose component are rotated by an arbitrary angle $\beta$ to $M_x$ and $M_y$:

$$\vec{H}_{pump\,\beta} = \frac{\Lambda_p}{M_s^2 - M_z^2} \begin{pmatrix} M_x \cos(\beta) - M_y \sin(\beta) \\ M_x \sin(\beta) + M_y \cos(\beta) \\ 0 \end{pmatrix} \qquad (11)$$

In particular, $\vec{H}_{pump\,\beta}$ is equal to $\vec{H}_{pump\,e,h}$ for $\beta = \pm 90°$.

After deriving the general expression for $\vec{H}_{pump}$, we use Eq. 7 to express the TLS terms $\Lambda_h, \Lambda_e, \gamma_v, \gamma_c, \gamma_{inh}, \omega$ and $V_{12}$ using terms from the LLGS equation, $\vec{M}, \alpha, H_0, H_{SHE_{DC}}, H_{SHE_{AC}}, \vec{N}_{D\,mag}, \Lambda_p$ and $\beta$. The demagnetization tensor is $\vec{N}_{D\,mag}$ where $N_x + N_y + N_z = N_0$. In our simulations, we assume spin currents generated by the SHE in a normal-metal\ferromagnet (NM\FM) bilayer, so that $H_{SHE_{AC\backslash DC}} =$ $\frac{\hbar \Theta_{SHE} J_{C\,AC\backslash DC}}{2e M_s t_{FM}}$, where $\Theta_{SHE}$ is the SHA (spin Hall angle), $t_{FM}$ is the thickness of the FM layer and $J_{C\,AC\backslash DC}$ is the amplitude of the AC\DC charge current injected into the NM. The critical DC charge current required to excite the STT oscillator by SHE is given by $J_{C\,STT} \equiv (2e\alpha H_0 M_s t_{FM})/[\hbar\, \Theta_{SHE}(1+\alpha^2)]$. We use the general case of a damped magnetic system (with Gilbert damping $\alpha$) in the presence of an external field $H_0$, that is driven by the AC and DC STT excitations $H_{SHE_{AC}}\vec{S}_{AC}$ and $H_{SHE_{DC}}\vec{S}_{DC}$, demagnetizing field $\vec{H}_{Dmag} = \sum_{i=x,y,z} N_i M_i \hat{\imath}$, and the $\vec{H}_{pump[\beta]}$ field presented in Eq. 11. We set the orientation of the fields and STT terms as following (Fig.1):

$$\vec{S}_{DC} = \hat{z}, \vec{S}_{AC} = \hat{x}$$

$$\vec{H}_{eff} = \vec{H}_0 + \vec{H}_{pump\,\beta} + \vec{H}_{Dmag} = \begin{pmatrix} \Lambda_p \frac{M_x \cos(\beta) - M_y \sin(\beta)}{M_s^2 - M_z^2} + N_x M_x \\ \Lambda_p \frac{M_x \sin(\beta) + M_y \cos(\beta)}{M_s^2 - M_z^2} + N_y M_y \\ H_0 + N_z M_z \end{pmatrix}$$

The comparison is summarized in Table 1. For the full mathematical derivation, see Appendix 2. From Table 1, we see that the $\hat{z}$ component of $\vec{H}_{Dmag}$ contributes directly to the external field, as seen in the terms $\gamma_v$, $\gamma_c$, $\gamma_{inh}$ and $\omega$. It enhances the Larmor frequency accordingly. The transverse $(\hat{x}, \hat{y})$ components of the demagnetization field were shown in Ref. [9] to act as additional oscillatory fields, which manifests in their contribution to $V_{12}$ in addition to the driving $H_{SHE_{AC}}$ excitation.

The $H_{SHE_{AC}}$ term changes $\omega$ in the periodicity of the precession (because of the $M_y$ term). In the presence of Gilbert damping and unequal transverse demagnetization ($N_x \neq N_y$), the transverse components of the demagnetization field change $\omega$ in double the periodicity of the precession (because of the $M_x M_y$ term). The explicit contribution of $\vec{H}_{pump}$ to $\omega$ is $\frac{\gamma \Lambda_p}{(1+\alpha^2)(M_s^2 - M_z^2)} \{\alpha M_s \sin(\beta) - M_z \cos(\beta)\}$, where the first term in the bracket is symmetric and the second is antisymmetric with respect to $M_z$. For $\beta = \pm 90°$ and in the presence of Gilbert damping, $\vec{H}_{pump\,e\backslash h}$ makes $\omega$ diverge near the poles of the Bloch sphere. However, around $\beta = 0°, 180°$ where $\vec{H}_{pump\,\beta}$ doesn't correspond to $\vec{H}_{pump\,e\backslash h}$, the change in $\omega$ is more pronounced as it doesn't depend on the small value of $\alpha$ anymore. This might be a possible explanation for the switching of a magnetic moment using optical (THz) frequencies in L \ R polarized optical pulses, since optical frequencies are 2-3 orders of magnitude higher than usual RF resonance frequencies used in conventional magnetization dynamics.

In last section of this paper, we show how $\vec{H}_{pump}$ can be used to switch $\vec{M}$. We first determine the condition at which the injection pump exceeds the damping in the equation of motion of $\dot{\rho}_{22}$ in Eq. 7:

$$\Lambda_e > \gamma_c \rho_{22} + [M_y \Re\{V_{12}\} + M_x \Im\{V_{12}\}] \qquad (12)$$

We insert $\Lambda_e, \gamma_c, \rho_{22}, \Im\{V_{12}\}$ from Table 1 and $H_{SHE_{AC}}$ (not including demagnetization terms), for $\beta = 90°$ (i.e. $\vec{H}_{pump\,\beta} = \vec{H}_{pump\,e}$):

$$|\Lambda_p| > \alpha \left(\frac{M_s^2 - M_z^2}{M_s}\right)\left[H_0\left(1 - \frac{J_{C\,DC}}{J_{C\,STT}}\right)\right] + (1 + \alpha^2)\left[\frac{\hbar\Theta_{SHE}J_{C\,AC}}{2et_{FM}}\frac{M_z M_x}{M_s^2}\right] \qquad (13)$$

In general, $M_x$ and $M_z$ depend on the opening cone angle ($\theta_{cone}$) of $\vec{M}$. Since $M_x, M_z \leq M_s$ the upper limit of $\frac{M_z M_x}{M_s^2}$ and $\frac{M_s^2 - M_z^2}{M_s^2}$ is 1. Using this upper limit, we obtain the condition:

$$|\Lambda_p| > \alpha M_s \left[H_0\left(1 - \frac{J_{C\,DC}}{J_{C\,STT}}\right)\right] + (1 + \alpha^2)\left[\frac{\hbar\Theta_{SHE}J_{C\,AC}}{2et_{FM}}\right] \qquad (14)$$

When this condition is fulfilled, $\theta_{cone} = 180°$ at steady state. From Eq. 14, the switching condition depends on both the AC and DC spin current, as well as on the external field. This dependence is expressed in Fig. 2(a) & (b), which stimulate $\theta_{cone}$ at steady state under various $|\Lambda_p|, J_{C\,AC}$ and $J_{C\,DC}$ values. In Fig. 2(a), $\theta_{cone}$ is depicted as a logarithmic function of $|\Lambda_p|$ for three $J_{C\,AC}$ values. By applying a DC current of the value $J_{C\,DC} = 0.8 \cdot J_{C\,STT}$ (dotted lines) the switching condition shifts to a lower $|\Lambda_p|$ value, according to the first term in Eq. 13. In Fig. 2(b) $\theta_{cone}$ is depicted as a logarithmic function of $J_{C\,AC}$, for $|\Lambda_p|$ values around the critical $\log_{10}(|\Lambda_p|) = 7.8\,[A^2/m^2]$ value (black dashed box in Fig. 2(a)), and for $J_{C\,DC} = 0$. In both Fig. 2(a) &(b), all the traces in which the switching condition is reached have a sharp behavior around their respective critical values, corresponding to a distinct switching condition for each trace. In Fig. 2(c) the temporal traces of $M_z$ and $(\vec{H}_{pump})_x$ are plotted. In this Figure, $\vec{H}_{pump}$ changes sign each time the switching condition defined by $|M_z| = 0.99M_s$ is met, effectively shifting between $\vec{H}_{pump[e]}$ (Eq. 10) and $\vec{H}_{pump[h]}$ (Eq. 9). The Lorentzian envelope of $\vec{H}_{pump}$ is especially noticeable near the switching condition.

The duration of a down-switch of $\vec{M}$ (in the $-\hat{z}$ direction) is much longer than for an up-switch of $\vec{M}$ (in the $\hat{z}$ direction). This behavior is explained by examining the torques acting on $\vec{M}$ in each case: when $\vec{H}_{pump}$ switches $\vec{M}$ downwards (to the higher Zeeman energy state), it has to first overcome the damping torque induced by $\vec{H}_0$, thus the effective torque switching $\vec{M}$ downwards is just a part of the torque induced by $\vec{H}_{pump}$. On the other hand, when $\vec{H}_{pump}$ switches $\vec{M}$ upwards, its torque adds to the damping torque induced by $\vec{H}_0$, thus the effective torque switching $\vec{M}$ upwards is larger than the torque induced by $H_{pump}$ alone. Lastly, in Fig. 2(d) we plot $\theta_{cone}$ at steady state under the influence of a general $\vec{H}_{pump\ \beta}$ whose components are rotated by an angle $\beta$ to $M_x$ and $M_y$. In particular, when $\beta = 90°$ (270°), $\vec{H}_{pump\ \beta}$ is equal to $\vec{H}_{pump\ e}$ ($\vec{H}_{pump\ h}$), and $\theta_{cone} = 180°$ (0°). At intermediate $\beta$ values, such as $\beta = 45°$, the response goes through Rabi oscillations before reaching the steady precession state. The parameters were chosen larger than for the previous calculations, explicitly $J_{C\ AC} = 6 \cdot 10^9 [A/m^2]$ , $J_{C\ DC} = 0 - 2 \cdot J_{C\ STT}$ and $\log_{10}(|\Lambda_p|) = 8 [A^2/m^2]$, since large torques were required in order to observe switching for all $\beta$ values. The Figure shows that the larger $J_{C\ DC}$ gets, the larger the region of $\beta$ values that are able to switch $\vec{M}$ downwards gets, and that the overall behavior sharpens when the switching condition is met in each trace - all in accordance with Eq. 14.

**Summary**

In summary, we present a detailed and rigorous model for magnetization dynamics in the density matrix picture. We explored the different excitations in the LLGS equation as terms in the density matrix formalism. We derived the field that is analogue to an optical pump. We demonstrated the effect of the demagnetization field on the Larmor frequency and AC interaction. We explained how the damping terms of the density matrix are modified by DC spin current and related to the STT oscillator phenomena. The implication of this work is to bring into light a more complete picture of the density matrix dynamics, as in the optical domain, damping is introduced only phenomenologically. In the LLGS picture, the arise from a physical Rayleigh friction process.

## Table 1

| TLS term | Magnetization Bloch vector | Role |
|---|---|---|
| $\gamma_v$ | $-\dfrac{\gamma\alpha}{M_s(1+\alpha^2)}\left\{\left[H_0\left(1-\dfrac{J_{C\,DC}}{J_{C\,STT}}\right)+M_zN_z\right](M_s-M_z)-\Lambda_p\cos(\beta)\right\}$ | The decay rate of the valence band |
| $\gamma_c$ | $\dfrac{\gamma\alpha}{M_s(1+\alpha^2)}\left\{\left[H_0\left(1-\dfrac{J_{C\,DC}}{J_{C\,STT}}\right)+M_zN_z\right](M_s+M_z)+\Lambda_p\cos(\beta)\right\}$ | The decay rate of the conductance band |
| $V_{12}$ | $\dfrac{\gamma}{2M_s}\left[jH_{SHE_{AC}}M_z+\dfrac{M_s-j\alpha M_z}{(1+\alpha^2)}(N_xM_x-jN_yM_y)\right]$ | AC excitation term |
| $\Lambda_h$ | $\dfrac{\gamma}{1+\alpha^2}\Lambda_p(\sin(\beta)-\alpha\cos(\beta))$ | Incoherent pump rate of holes injected to the valence band |
| $\Lambda_e$ | $\dfrac{\gamma}{1+\alpha^2}\Lambda_p(\sin(\beta)+\alpha\cos(\beta))$ | Incoherent pump rate of electrons injected to the conduction band |
| $\omega$ | $\dfrac{\gamma}{(1+\alpha^2)}\left\{H_0+M_zN_z+H_{SHE_{AC}}(1+\alpha^2)\dfrac{M_y}{M_s}\right.$ $-\Lambda_p\dfrac{M_z\cos(\beta)}{M_s^2-M_z^2}$ $+\dfrac{\alpha}{M_s}\left[\Lambda_p\sin(\beta)\left(\dfrac{M_s^2}{M_s^2-M_z^2}\right)\right.$ $\left.\left.-(N_x-N_y)M_xM_y\right]\right\}$ | The resonance frequency, with a slight correction to the Larmor frequency |
| $\gamma_{inh}$ | $\dfrac{\gamma}{(1+\alpha^2)}\dfrac{M_z}{M_s}\left\{\Lambda_p\dfrac{\alpha M_z\cos(\beta)-M_s\sin(\beta)}{M_s^2-M_z^2}\right.$ $\left.+\alpha\left[H_0\left(1-\dfrac{J_{C\,DC}}{J_{C\,STT}}\right)+M_zN_z\right]\right\}$ | The decay rate term due to inhomogeneous broadening |

**Figure 1**

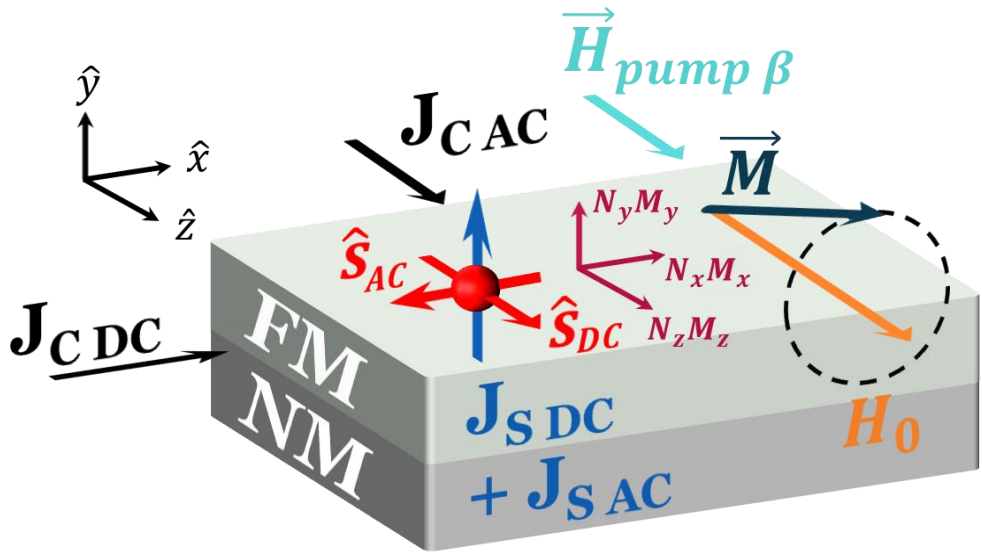

**Fig. 1. NM\FM bilayer.** Illustration of the excitations acting on the magnetization in this model – external field, AC and DC STT, demagnetization field and the $H_{pump}$ field derived in this paper.

**Figure 2**

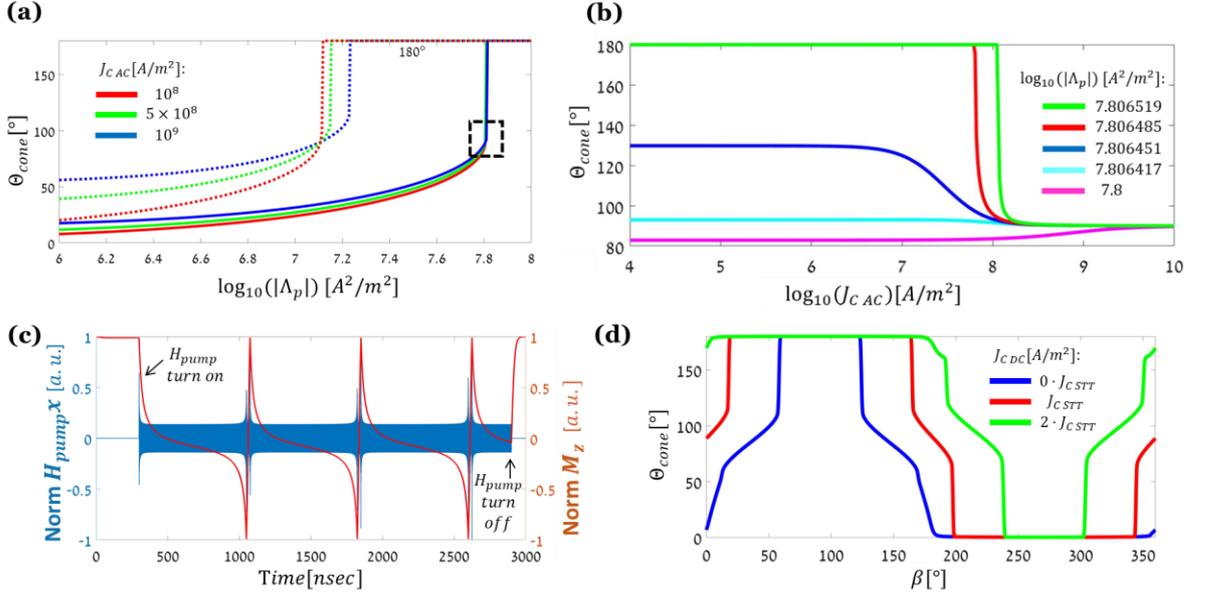

Fig. 2. Magnetization switching schemes using $H_{pump}$. (a) Opening cone angle at steady state as a function of $|\Lambda_p|$ for three $J_{C\,AC}$ values. The dashed lines correspond to $J_{C\,DC} = 0.8 \cdot J_{C\,STT}$ and the solid lines correspond to $J_{C\,DC} = 0$. (b) Opening cone angle at steady state as a function of $J_{C\,AC}$, for five $|\Lambda_p|$ values around the critical $log_{10}(|\Lambda_p|) = 7.8 [A^2/m^2]$ value, at $J_{C\,DC} = 0$. This region corresponds to the black dashed box in Fig. 2(a). (c) Temporal traces of the normalized $M_z$ (red) and $H_{pump\,x}$ (blue) at $J_{C\,DC} = 0$, $J_{C\,AC} = 6 \cdot 10^8 [A/m^2]$ and $log_{10}(|\Lambda_p|) = 7.8118 [A^2/m^2]$. $H_{pump}$ changes sign after each switching of $\vec{M}$. (d) The opening cone angle at steady state as a function of $\beta$ for different $J_{C\,DC}$ values, $J_{C\,AC} = 6 \cdot 10^9 [A/m^2]$ and $log_{10}(|\Lambda_p|) = 8 [A^2/m^2]$.